\documentclass[twoside]{dis09}
\usepackage[latin1]{inputenc}
\usepackage[dvips]{graphicx,epsfig,color}
\usepackage{wrapfig,rotating}
\usepackage{amssymb,amsmath,array}

\newcommand{\beq}{\begin{eqnarray}}
\newcommand{\eeq}{\end{eqnarray}}
\newcommand{\la}{\langle}
\newcommand{\ra}{\rangle}
\def\xhat{\widehat{x}}
\def\zhat{\widehat{z}}

\begin{document}
\title{Theoretical Update of Twist-3 Single-Spin Asymmetry in Semi-Inclusive DIS}

\author{Yuji Koike$^1$ and Kazuhiro Tanaka$^2$
%
%
\vspace{.3cm}\\
%
1- Department of Physics, Niigata University,
Ikarashi, Niigata 950-2181, Japan
%
\vspace{.1cm}\\
2- Department of Physics, Juntendo University, Inba, Chiba 270-1695, Japan}

\maketitle

\begin{abstract}
We discuss
the single-spin asymmetry in semi-inclusive DIS, 
$ep^\uparrow\to e\pi X$, based on the twist-3 mechanism in the
collinear factorization relevant for the pion production with
the large transverse-momentum.
This updates our previous study by including, in particular, the contributions 
induced by the novel partonic subprocesses, 
and allows us to derive the entire formula for
the corresponding single-spin asymmetry 
associated with
a complete set of the twist-3 quark-gluon correlation functions 
in the transversely polarized nucleon.
We discuss the correspondence with the results obtained by the transverse-momentum-dependent 
factorization relevant for the pion production
with the low transverse-momentum.
\end{abstract}

The single transverse-spin asymmetry (SSA) in the semi-inclusive DIS (SIDIS), 
$e(\ell)+p(p,S_\perp)\to e(\ell')+\pi(P_h)+X$,
is observed as {\em T-odd} effect 
in the cross section for the scattering of transversely polarized nucleon with momentum $p$
and spin $S_\perp$,
off unpolarized lepton
with momentum $\ell$, 
producing a pion with momentum $P_h$ 
which is observed in the final state. Here, $q=\ell-\ell'$, $Q^2=-q^2$,
and $Q \gg \Lambda_{\rm QCD}$.
The SSA can be observed also in $pp$ collisions,
the pion production $p^\uparrow p\to\pi X$~\cite{To},
and the Drell-Yan and direct-$\gamma$ production, 
$p^\uparrow p\to \gamma^{(*)} X$~\cite{JQVY06DY}.
Similarly as 
these 
examples, 
the SSA in the SIDIS requires,
(i) nonzero $P_{h\perp}$ originating 
from transverse motion
of quark or gluon; 
(ii) nucleon helicity flip; 
and (iii) interaction beyond Born level to produce the
interfering phase for the cross section. 
When $P_{h\perp}\ll Q$, all (i)-(iii) may be generated nonperturbatively
from 
the T-odd, transverse-momentum-dependent (TMD) parton distribution/fragmentation
functions 
such as the Sivers function (see~\cite{Bacchetta:2006tn,JQVY06SIDIS}).
By contrast,
for large $P_{h\perp}\gg \Lambda_{\rm QCD}$,
(i) should come from perturbative mechanism
as the recoil from the hard (unobserved) final-state partons, 
while nonperturbative effects 
can participate in the other two, (ii) and (iii), allowing us to obtain large SSA.
This 
is realized 
with the twist-3 distribution/fragmentation functions
in 
the collinear-factorization framework.
In our recent paper~\cite{EKT07},  
we have derived the corresponding factorization formula,
in the leading-order 
perturbative QCD,
for the SSA
associated with the twist-3 
distributions for the nucleon,
and provided a practical procedure to calculate the 
relevant partonic hard parts
manifesting their gauge invariance at the twist-3 level.
The factorization formula in~\cite{EKT07}
implies, for the 
form of the twist-3 single-spin-dependent
cross section 
as a function of
the azimuthal angle $\phi_h$ of the final-state pion~($\Phi\equiv \phi_h-\phi_S$)~\cite{KT071},
\begin{equation}
\frac{d^5\sigma^{\rm tw3}}{[d\omega]}
= 
\left( \sigma_1^{\rm tw3}
+\sigma_2^{\rm tw3}\cos\phi_h
+\sigma_3^{\rm tw3}\cos 2\phi_h \right) \sin\Phi
+ 
\left( \sigma_4^{\rm tw3}\sin\phi_h
+\sigma_5^{\rm tw3}\sin 2\phi_h \right) \cos\Phi\ ,
\label{tw3}
\end{equation}
in a frame where the 3-momenta $\vec{q}$ and $\vec{p}$ of the
virtual photon and the transversely polarized nucleon 
are collinear along the $z$ axis;
$\phi_h$ as well as the azimuthal angle $\phi_S$ 
of the nucleon's transverse-spin vector $S_\perp^\mu$
is measured 
from the lepton plane,
and 
$[d\omega]\equiv dx_{bj}dQ^2 dz_f dq_T^2 d\phi_h$ denotes
the differential element
with the usual kinematical variables for the SIDIS, 
$x_{bj}={Q^2/ (2p\cdot q)}$, $z_f={p\cdot P_h / p\cdot q }$, and
$q_T = P_{h\perp}/z_f$.
In~\cite{EKT07}, we calculated the explicit form 
only for $\sigma^{\rm tw3}_{1,2,3}$. 
In addition, 
a new type of partonic subprocesses
that contribute to the twist-3 mechanism has been pointed out in~\cite{KVY08}.
In this report we present the 
main features of the full result~\cite{KT09}, taking into account these points 
as well as yet another type of the novel partonic subprocesses relevant to
the twist-3 mechanism.

In the twist-3 mechanism relevant to (\ref{tw3}), 
the nucleon helicity flip of (ii) above 
is provided by the quark-gluon correlation
in its inside,
which generates additional nonperturbative gluon 
as described by
the nucleon matrix element of the
three-body nonlocal operators on the lightcone,
$\la p\ S_\perp|\bar{\psi}(0)F^{\alpha +}(\xi n) \psi(\lambda n) |p\ S_\perp\ra$, where
$n^2 =0$, $p\cdot n=1$, and $F^{\alpha \beta}$ is the gluon field strength tensor. 
The Fourier transform of this matrix element defines
the two 
twist-3 correlation
functions, $G_F(x_1,x_2)$ and $\widetilde{G}_F(x_1,x_2)$, as the 
coefficients of the independent Lorentz structures,
where $x_{1,2}$ denote the lightcone 
momentum fractions associated with the quark fields $\psi, \bar{\psi}$;
$G_F$ and $\widetilde{G}_F$ are symmetric and antisymmetric, respectively, 
under $x_1\leftrightarrow x_2$
and form a complete set to calculate (\ref{tw3})~\cite{EKT06}.
The above additional gluon
from the transversely polarized nucleon,
carrying the momentum fraction $x_2 -x_1$,
participates
into the partonic hard scattering, as in the diagrams in Figs.~1, 2 below.
The coupling of this gluon allows
an internal propagator in the partonic subprocess to be on-shell,
and this produces the imaginary phase of (iii) above, as the pole contribution using
$1/(k^2 + i\varepsilon) = {\rm P}(1/k^2) - i\pi\delta(k^2)$.
Depending on the resulting value of the momentum fractions $x_{1,2}$ at such poles, 
these poles are the soft-gluon pole (SGP) for $x_1=x_2$, the soft-fermion pole (SFP)
for $x_i=0$,
and the hard pole (HP) for $x_i = x_{bj}$, where $i=1$ or~2.
We note that the right four diagrams of Fig.~1 
reduce to those representing the partonic hard part
of the twist-2 unpolarized cross section 
for the large-$P_{h\perp}$ pion production, $ep \to e\pi X$, if
we remove the insertion of the additional gluon.
In~\cite{EKT07}, we calculated all the partonic subprocesses 
associated with such {\em coherent}
gluon. By contrast, the diagrams that are not associated with the coherent
gluon, like the left four diagrams of Fig.~1 and the diagrams in Fig.~2,
represent the novel partonic subprocesses~\cite{KVY08,KT09}; 
these contributions
do not have straightforward counterpart in the TMD factorization approach,
i.e., 
in the effects due to 
the intrinsic transverse motion of the valence Fock components
in the nucleon.
Now 
we summarize the main points for updating
each of the SGP, SFP and HP contributions,
compared with our previous results~\cite{EKT07}.

First of all, 
the SGP contribution occurs only
from the diagrams involving the coherent gluon 
(see Figs.~8 and~10 of~\cite{EKT07}). 
Thus, our task here
is to calculate the SGP contribution to $\sigma^{\rm tw3}_{4,5}$ of (\ref{tw3}) from the
relevant diagrams,
following the procedure given in~\cite{EKT07}.   
Alternatively, 
one can use 
the master formula proved in~\cite{KT071} (see also~\cite{KT072}),
yielding
the full SGP contribution to (\ref{tw3}) as
\begin{equation}
\frac{d^5 \sigma_{\rm SGP}^{\rm tw3}}{[d\omega]} 
= 
\frac{\pi M_N}{2 C_F z_f}
\left(
\sin\Phi
\frac{\partial}{\partial q_T}
+ 
\frac{\cos\Phi}{q_T}\frac{\partial}{\partial \phi_h}
\right)
\left. \frac{d^5\sigma_{\rm unpol}^{\rm tw2}}{[d\omega]}
\right|_{f_q(x)\rightarrow G_F^q(x,x),\ D_j(z) \rightarrow {\cal C}_j zD_j (z)} ,
\label{sgp}
\end{equation}
where $C_F=(N_c^2 -1)/(2N_c)$, $M_N$ is the nucleon mass, and 
(${\cal N}_q \equiv \alpha_{em}^2 \alpha_s e_q^2 / (8\pi x_{bj}^2 S_{ep}^2 Q^2)$)
\begin{equation}
\frac{d^5\sigma_{\rm unpol}^{\rm tw2}}{[d\omega]}
=\!\! \sum_{j=q,g} {\cal N}_q  \!
\int\! \frac{dz}{z} \int\! \frac{dx}{x} 
D_j(z) f_q(x)
\sum_{n=1}^3 \widehat{\varrho}^{jq}_n \cos\left((n-1)\phi_h \right) 
\delta \! \left( {q_T^2\over Q^2} -
\left[ {1\over \xhat} -1\right]\left[{1\over \zhat}-1\right]\right)
\label{tw2}
\end{equation}
is the twist-2 unpolarized cross section 
for $ep \rightarrow e\pi X$,
obeying the factorization formula
with
the twist-2, parton-fragmentation and quark-distribution 
functions, $D_j(z)$ and $f_q(x)$, 
for the pion and unpolarized nucleon, respectively.
Here, $S_{ep}=(\ell+p)^2$, $\xhat=x_{bj}/x$, $\zhat=z_f/z$, and
the sum over all quark and antiquark flavors,
$q=u, \bar{u}, d, \bar{d}, \cdots$,
is implicit for 
the index $q$;
$\widehat{\varrho}_n^{jq}$ 
denote the 
partonic hard parts 
and read
$\widehat{\varrho}_1^{jq}=  (1+\cosh^2\psi)\widehat{\sigma}_1^{jq}-2\widehat{\sigma}_2^{jq}$,
$\widehat{\varrho}_2^{jq}= -\widehat{\sigma}_3^{jq} \sinh 2\psi$, and
$\widehat{\varrho}_3^{jq}= \widehat{\sigma}_4^{jq}\sinh^2\psi$, 
where $\cosh\psi = 2x_{bj}S_{ep}/ Q^2 -1$, and $\widehat{\sigma}_k^{jq}$ 
are 
given in Eqs.~(57), (59) of~\cite{EKT07} as the functions of $\xhat$, $\zhat$, $q_T$, and $Q$.
For the replacements indicated in (\ref{sgp}),
${\cal C}_{q}=-1/(2N_c)$, ${\cal C}_g=N_c/2$, and
$G_F^q(x,x)$ for the flavor $q$ arises at the SGP corresponding to $x_1=x_2$ while
$\widetilde{G}_F^q(x,x)=0$ due to the 
symmetry property. 
The $\phi_h$-dependence of (\ref{tw2}) resides in $\cos\left((n-1)\phi_h \right)$
only,  
so that
$\sigma^{\rm tw3}_{4}$ 
and $\sigma_5^{\rm tw3}$ of (\ref{tw3})
are generated by the angular derivative in (\ref{sgp})
and are related to
$\sigma_2^{\rm tw3}$ and $\sigma_3^{\rm tw3}$, respectively.
However, $\sigma^{\rm tw3}_{4,5}$ do not receive
the derivative  
of the twist-3 distribution, $dG_F^q(x,x) /dx$, 
which contributes to $\sigma^{\rm tw3}_{1,2,3}$ through the $q_T$-derivative in (\ref{sgp})
(see Erratum of~\cite{KT071} for the explicit formula obtained from (\ref{sgp})).

Second, 
the SFP 
corresponding to $x_1 =0$
in the diagrams of Fig.~1,
the right four diagrams 
involving 
the coherent gluon~\cite{EKT07} and the others of new type~\cite{KVY08}, 
picks up $G_F^q(0,x)$ as well as $\widetilde{G}_F^q(0,x)$.
As pointed out in \cite{KVY08}, the contributions of
these two types of diagrams cancel with each other 
when the final-state parton with the flavor $q$, represented by
the upper horizontal line across the cut, fragments into the pion.
Thus, the SFP contributes to $\sigma^{\rm tw3}_{1,\cdots,5}$ of (\ref{tw3})
accompanying $D_{j}(z)G_F^q(0,x)$ or $D_{j}(z) \widetilde{G}_F^q(0,x)$ 
with $j=\bar{q}, g$, corresponding to the case where the lower horizontal line
across the final-state cut 
in Fig.~1 represents the fragmentation function $D_{j}(z)$.
\begin{figure}
\centerline{\includegraphics[height=.24\textheight]{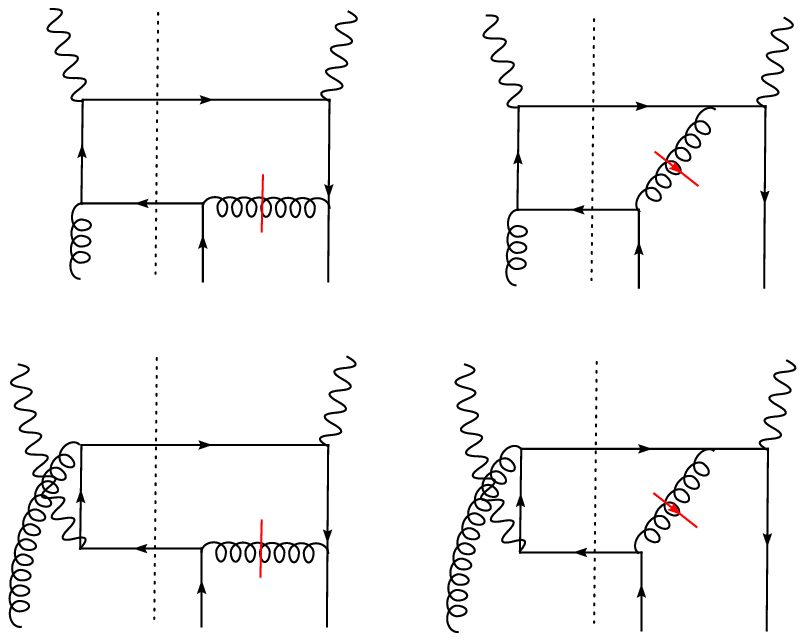}
\hspace{1.0cm}
\includegraphics[height=.24\textheight]{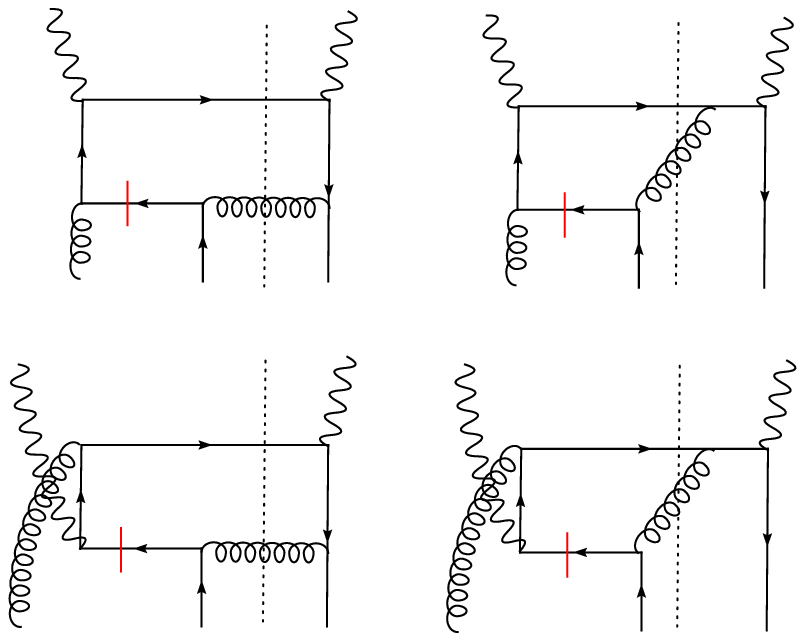}}
  \caption{Diagrams for the SFP contribution to SSA in SIDIS.
The bar denotes 
the pole contribution of the propagator. Either of the two partons 
going across the cut,
the dotted line, 
fragments into the pion, while the other is unobserved.
Mirror diagrams also contribute.}
\end{figure}
Calculating these relevant diagrams with the procedure in~\cite{EKT07},
our result obeys a simple relation
that, for each of $\sigma^{\rm tw3}_{1,\cdots,5}$ in (\ref{tw3}), 
the partonic hard parts associated with $D_{j}(z)G_F^q(0,x)$ 
differ only in the overall sign between $j=\bar{q}$ and $j=g$, 
and similarly for those associated with $D_{j}(z) \widetilde{G}_F^q(0,x)$; this relation reflects 
the above-mentioned cancellation for the channel corresponding to $D_{q}(z)G_F^q(0,x)$,
$D_{q}(z) \widetilde{G}_F^q(0,x)$.

Third,
the HP corresponding to $x_1 = x_{bj}$
in the diagrams involving the coherent gluon (see Fig.~2 in~\cite{EKT07})
can be evaluated according to~\cite{EKT07},
and picks up $D_j(z) G_F^q (x_{bj}, x)$ and $D_j(z) \widetilde{G}_F^q (x_{bj}, x)$
with $j=q,g$.
Decomposing the result into the independent azimuthal structures 
as in (\ref{tw3}),
we find the new results for $\sigma^{\rm tw3}_{4,5}$, in addition
to $\sigma^{\rm tw3}_{1,2,3}$ obtained in~\cite{EKT07}.
The results show, for each term associated with 
$D_j(z) G_F^q (x_{bj}, x)$, $D_j(z) \widetilde{G}_F^q (x_{bj}, x)$, that 
$\sigma_4^{\rm tw3}$ and $\sigma_5^{\rm tw3}$ coincide with 
$\sigma_2^{\rm tw3}$ and $\sigma_3^{\rm tw3}$, respectively.
However, in addition to those diagrams 
associated with the $qg$ final states,
the 
partonic subprocesses with the $q\bar{q}$ final states
can also receive the HP contribution at $x_2 = x_{bj}$, as illustrated in Fig.~2.
\begin{figure}
\centerline{\includegraphics[height=.12\textheight,clip]{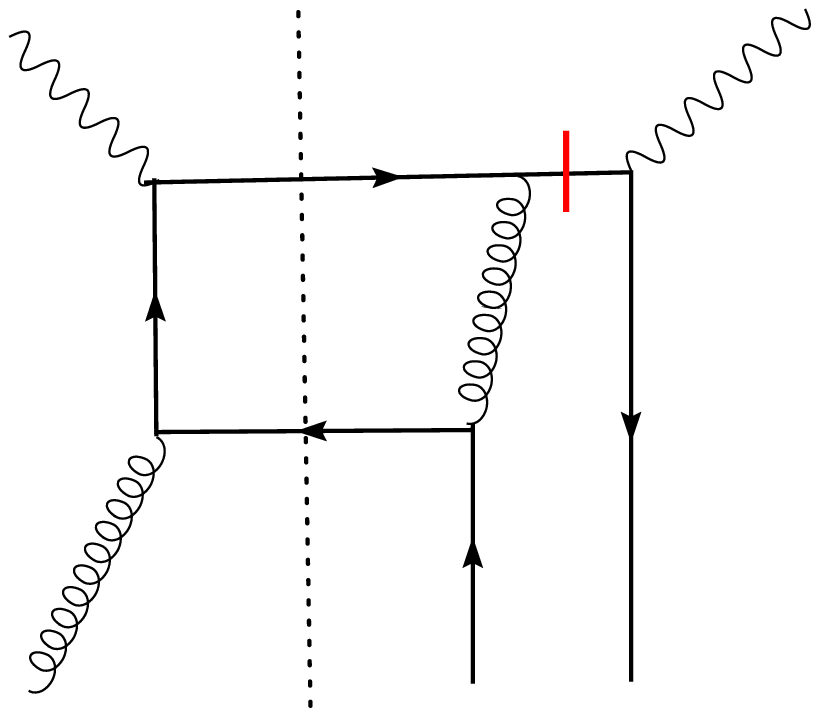}\hspace{0.4cm}
\includegraphics[height=.12\textheight,clip]{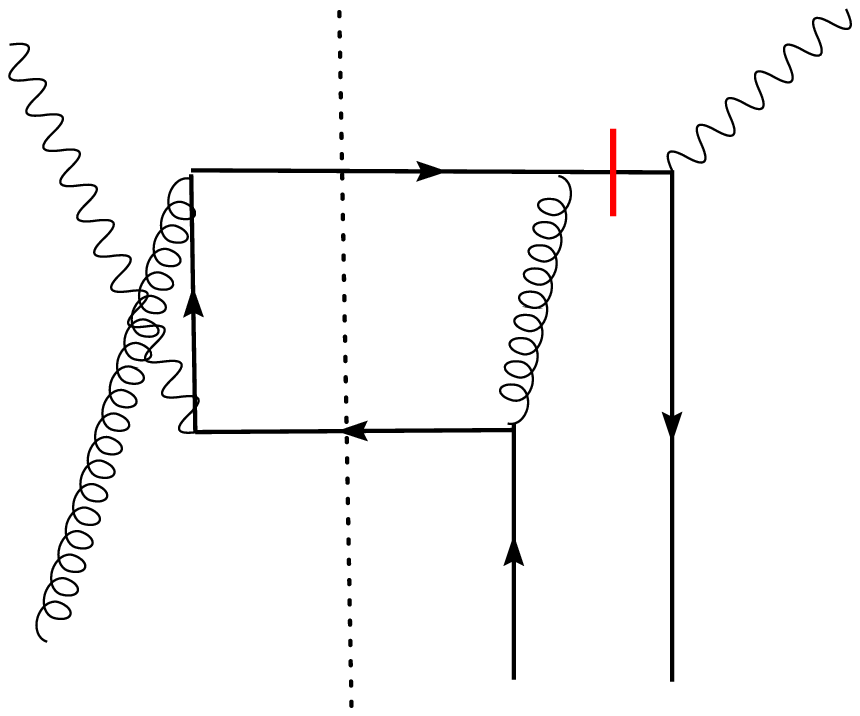}\hspace{0.4cm}
\includegraphics[height=.12\textheight,clip]{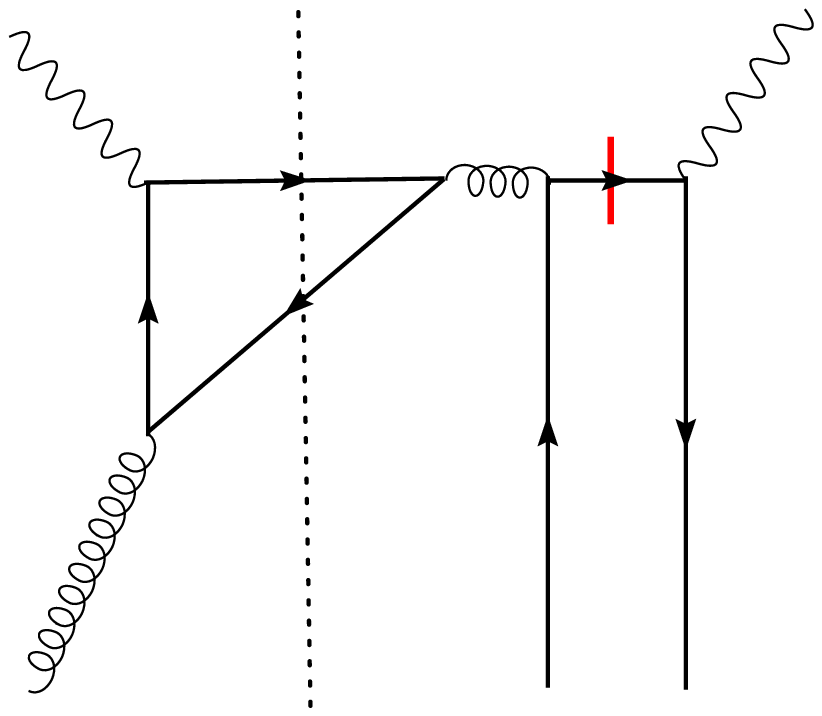}\hspace{0.4cm}
\includegraphics[height=.12\textheight,clip]{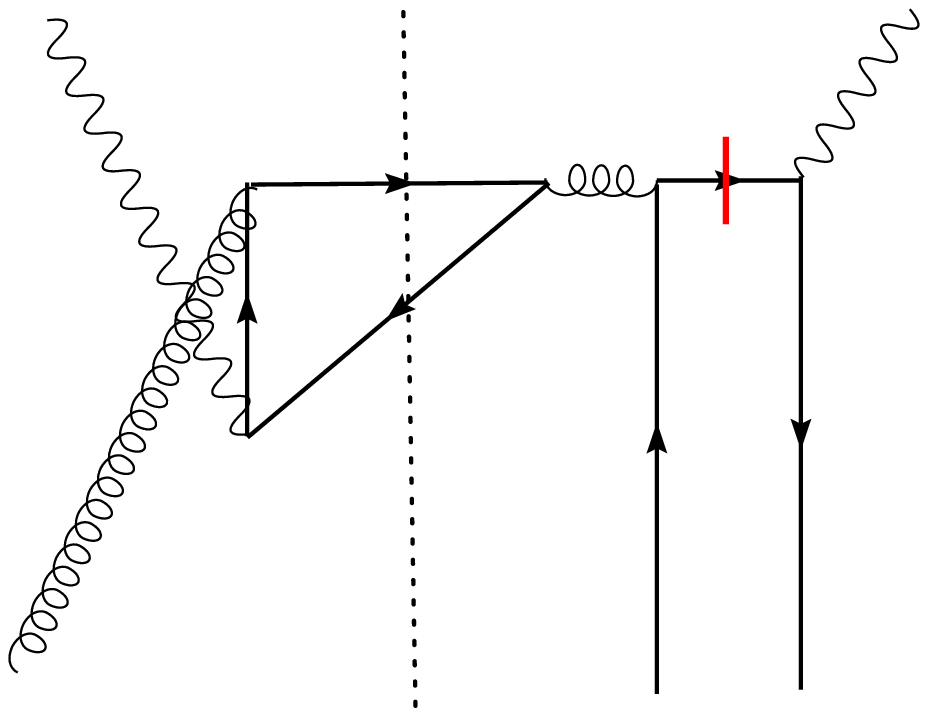}}
\caption{Same as Fig.~1, but for the new subprocesses that give rise to the HP contribution.}
\end{figure}
This would seem formally similar to the situation in the SFP contribution of Fig.~1,
but, by contrast, the HP contribution of the novel diagrams of Fig.~2
does not have simple relation, for any fragmentation channel, to 
the above-mentioned HP contribution induced by the coherent gluon.
Indeed, the HP in Fig.~2
picks up 
$D_j(z) G_F^q (x_{bj}, x_{bj}-x)$ and $D_j(z) \widetilde{G}_F^q (x_{bj}, x_{bj}-x)$
with $j=q, \bar{q}$, where $x_{bj}-x <0$ 
indicating that the nucleon absorbs a coherent $q\bar{q}$ pair 
in the RHS of the final-state cut in Fig.~2.
Calculating these novel HP diagrams with the procedure of~\cite{EKT07},
we find $\sigma_2^{\rm tw3,HP} \neq \sigma_4^{\rm tw3,HP}$
and $\sigma_3^{\rm tw3,HP} \neq \sigma_5^{\rm tw3,HP}$ 
for the azimuthal structures of (\ref{tw3}).

Collecting all the pole contributions discussed above,
one obtains $\sigma^{\rm tw3}_n$ 
in the 
single-spin-dependent 
cross section (\ref{tw3}),
associated with the twist-3 
functions
$G_F$ and $\widetilde{G}_F$, 
as
\beq
\lefteqn{\!\!\!\!\!\! \sigma_n^{\rm tw3}
=
\frac{-\pi M_N {\cal N}_q}{4}\!\!
\sum_{j=q,\bar{q},g}
\int \frac{dz}{z} \int \frac{dx}{x}
D_j(z)\left[
\widehat{\sigma}^{jq}_{Dn} x \frac{dG_F^q(x,x)}{dx} 
+\widehat{\sigma}^{jq}_{Gn} G_F^q(x,x) + \widehat{\sigma}^{jq}_{Fn} G_F^{q}(0,x) 
\right.} \nonumber \\
&& \!\!\!\!\!\!\!\!\!\!\!\!\!\!\!\!
\left. +\
\widehat{\sigma}^{jq}_{HOn}
G_F^q(x_{bj},x)
+ \widehat{\sigma}^{jq}_{HNn} G_F^q(x_{bj},x_{bj}-x)
\right]
\delta\! \left(\frac{q_T^2}{Q^2}-\left[1-\frac{1}{\widehat{x}}\right]\left[
1-\frac{1}{\widehat{z}}\right]\right)\! +\! 
\left[\widetilde{G}_F \mbox{ terms} \right] ,
\label{tw3n}
\eeq
for $n=1,2,3,4,5$, where the partonic hard parts obey
$\widehat{\sigma}^{\bar{q}q}_{Dn} =\widehat{\sigma}^{\bar{q}q}_{Gn} 
=\widehat{\sigma}^{qq}_{Fn}= \widehat{\sigma}^{\bar{q}q}_{HOn}=\widehat{\sigma}^{gq}_{HNn}=0$,
$\widehat{\sigma}^{\bar{q}q}_{Fn}=-\widehat{\sigma}^{gq}_{Fn}$,
$\widehat{\sigma}^{jq}_{HO2}=\widehat{\sigma}^{jq}_{HO4}$,
and $\widehat{\sigma}^{jq}_{HO3}=\widehat{\sigma}^{jq}_{HO5}$,
and ``[$\widetilde{G}_F$ terms]'' have the similar structure as the preceding terms
except that the SGP contributions vanish
($\widetilde{G}_F(x,x)=0$).
The results (\ref{sgp}), (\ref{tw2}) further tell us that 
$\widehat{\sigma}^{jq}_{Dn}=-[4{\cal C}_j q_T\widehat{x}/C_F Q^2
(1-\widehat{z})] \widehat{\varrho}^{jq}_{n}$ and
$\widehat{\sigma}^{jq}_{Gn}=-(4{\cal C}_j q_T/C_FQ^2)
\{ (Q^2/\widehat{z}) \partial  \widehat{\varrho}^{jq}_{n} /\partial q_T^2 -
[\widehat{x}/(1-\widehat{z})] \partial ( \widehat{x} \widehat{\varrho}^{jq}_{n}) 
/\partial \widehat{x} \}$, for $n=1,2,3$, while
$\widehat{\sigma}^{jq}_{Dn}=0$ and 
$\widehat{\sigma}^{jq}_{Gn}=(2^{n-3}{\cal C}_j /C_F q_T \widehat{z})
\widehat{\varrho}^{jq}_{n-2}$, for $n=4,5$.
The explicit form of $\widehat{\sigma}^{jq}_{Fn}$, $\widehat{\sigma}^{jq}_{HOn}$,
$\widehat{\sigma}^{jq}_{HNn}$
and of the corresponding partonic hard parts for the 
``[$\widetilde{G}_F$ terms]''
will be presented elsewhere~\cite{KT09}.
We note that the delta function in (\ref{tw3n}) vanishes
when $x < x_{bj}\{1+z_fq_T^2/[(1-z_f)Q^2]\}$ or $z < z_{f}\{1+x_{bj}q_T^2/[(1-x_{bj})Q^2]\}$.
We recast (\ref{tw3}) into
\beq
\frac{d^5\sigma^{\rm tw3}}{[d\omega]}
&=&\sin(\phi_h-\phi_S)\,F^{\sin(\phi_h-\phi_S)}
+\sin(2\phi_h-\phi_S)\,F^{\sin(2\phi_h-\phi_S)}
+\sin\phi_S \,F^{\sin \phi_S}\nonumber\\
&&+\sin(3\phi_h-\phi_S)\,F^{\sin(3\phi_h-\phi_S)}
+\sin(\phi_h+\phi_S)\,F^{\sin(\phi_h+\phi_S)}\ ,
\label{singlesign}
\eeq
with the structure functions for the independent azimuthal structures,
$F^{\sin(\phi_h-\phi_S)}=\sigma_1^{\rm tw3}$, 
$F^{\sin(2\phi_h-\phi_S)}=(\sigma_2^{\rm tw3}+\sigma_4^{\rm tw3})/2$, 
$F^{\sin \phi_S}=(\sigma_4^{\rm tw3} -\sigma_2^{\rm tw3})/2$, 
$F^{\sin(3\phi_h-\phi_S)}=(\sigma_3^{\rm tw3}+\sigma_5^{\rm tw3})/2$, and
$F^{\sin(\phi_h+\phi_S)}=(\sigma_5^{\rm tw3} -\sigma_3^{\rm tw3})/2$,
which are all determined by the 
two independent twist-3 distributions of the nucleon as (\ref{tw3n}).
By contrast, when the transverse momentum $P_{h\perp}(=z_f q_T)$ of the pion is low,
the relevant single-spin-dependent cross section is calculated by
the TMD factorization approach~\cite{Bacchetta:2006tn},
and 
the corresponding five structure functions defined similarly as (\ref{singlesign}) 
are expressed
by 
the (many) independent
TMD distributions of the nucleon.

Our results (\ref{tw3n}), (\ref{singlesign}) hold for 
$q_T \gg \Lambda_{\rm QCD}$, 
and their small-$q_T$ behavior with $q_T \ll Q$ allows us to
make further contact to  
the {\em high}-$q_T$ behavior of the TMD factorization result
with $\Lambda_{\rm QCD} \ll q_T \ll Q$.
The corresponding small-$q_T$ behavior of~(\ref{singlesign}) was 
studied in~\cite{Bacchetta:2008xw}, 
using our previous result~\cite{EKT07,EKT06} for $\sigma_n^{\rm tw3}$. 
With our present update~(\ref{tw3n}), 
we obtain the small-$q_T$ behavior,
$F^{\sin(\phi_h-\phi_S)}\sim 1/q_T^3$, 
$F^{\sin(2\phi_h-\phi_S)}\sim 1/q_T^2$, 
$F^{\sin \phi_S}\sim 1/q_T^2$, 
$F^{\sin(3\phi_h-\phi_S)}\sim 1/q_T$, and
$F^{\sin(\phi_h+\phi_S)}\sim 1/q_T$, 
which are actually controlled by the SGP and HP contributions because
the SFP contributions are suppressed by the powers of $q_T$
for all five structure functions.
Although these leading power behaviors are formally same as those of the previous 
results~\cite{Bacchetta:2008xw} mentioned above, 
now the novel HP contribution from the diagrams in Fig.~2 also 
contributes to these leading-power terms, except for $F^{\sin(3\phi_h-\phi_S)}$.
In this connection, it has been shown in~\cite{JQVY06SIDIS,KVY08} 
that the most dominant $\sim 1/q_T^3$ term
from $F^{\sin(\phi_h-\phi_S)}$ coincides with the 
high-$q_T$ behavior of the Sivers effect 
in the TMD factorization,
where, however, the novel HP contribution was not taken into account.
These facts call for clarifying the role of the novel HP contribution
concerning the connection between the twist-3 and Sivers mechanism,
as well as its quantitative relevance in the cross section~(\ref{singlesign}).
We also mention that (\ref{singlesign}) 
receives 
another twist-3 effect
associated with the twist-3 fragmentation function of the pion, $\widehat{E}_F(z_1, z_2)$, 
for which we need more study.

To summarize,
we have discussed the twist-3 mechanism for the SSA in SIDIS,
induced by the quark-gluon correlation inside the nucleon, and 
updated and completed
the formula for 
the corresponding 
single-spin-dependent cross section.
Our cross section is expressed by the five structure functions 
corresponding to independent azimuthal structures, and
also includes the novel partonic subprocesses associated 
with the $q\bar{q}$ final states, as well as the previously known 
subprocesses associated 
with 
the $qg$ final states.
The novel contributions are 
related to 
non-minimal parton configurations inside the nucleon, 
describing
interference 
between scattering from a coherent $q\bar{q}$ pair and from a single gluon.
We finally note that our twist-3 mechanism disappears for the inclusive DIS,
because of cancellation of the final-state interactions 
for the unobserved final states~\cite{KT071,KT09}.

\vspace{0.2cm}
We thank F.~Yuan and A.~Metz for useful discussions.


\begin{footnotesize}



%

\end{footnotesize}


\end{document}